# CONE CRUSHER MODEL IDENTIFICATION USING BLOCK-ORIENTED SYSTEMS WITH ORTHONORMAL BASIS FUNCTIONS

Oleksii Mykhailenko[1]

[1]Department of Power Supply and Energy Management, State institution of higher education «KryvyiRih National University», KryvyiRih, Ukraine

*ABSTRACT*

*In this paper, block-oriented systems with linear parts based on Laguerre functions is used to approximation of a cone crusher dynamics. Adaptive recursive least squares algorithm is used to identification of Laguerre model. Various structures of Hammerstein, Wiener, Hammerstein-Wiener models are tested and the MATLAB simulation results are compared. The mean square error is used for models validation.It has been found that Hammerstein-Wiener with orthonormal basis functions improves the quality of approximation plant dynamics. The mean square error for this model is 11% on average throughout the considered range of the external disturbances amplitude. The analysis also showed that Wiener model cannot provide sufficient approximation accuracy of the cone crusher dynamics. During the process it is unstable due to the high sensitivity to disturbances on the output.The Hammerstein-Wiener model will be used to the design nonlinear model predictive control application.*

## 1. INTRODUCTION

In order to create technological unit control systems of ore dressing production it is necessary to have the mathematical model of the unit, accurately describing its characteristics. Taking into account the nonlinear properties of technological processes [1] of ore dressing and ore preparation, it is impossible to use linear dynamic models, which are studied well enough and have wide mathematical tool.

Approximation of nonlinear system characteristics is carried out using models based on the Volterra series, neuro-fuzzy models or block-oriented models (Hammerstein, Wiener and Hammerstein-Wiener). In terms of control the preference is given to use of latest models that represent the series of nonlinear blocks in a certain order, described by static functions, and linear block that displays the dynamics of the plant [2, 3]. Such model configuration determines simplicity of its practical implementation.

In this work as a linear block the model that is based on the Laguerre orthonormal functions[4,5 , 6] is used. Such a choice is connected with some features,which the specified model has. For example, in case of model parametrical identification there is no need for aprior information about time delays and time constants. Due to the orthogonality property the sufficient alignment with plant characteristics is achieved at the small model order. Thus, coefficients of model are estimated by methods of the regression analysis.

This article consists of four sections. The second section provides a mathematical description of block-oriented systems taking into account the representation of their linear part by orthonormal functions. The third section is concerned with the efficiency analysis of the nonlinear model parametric identification by performing the computational experiment in MATLAB software package.

## 2. NONLINEAR MODELS

The plant dynamics can be described by the model built on the basis of Laguerre orthonormal functions, which indiscrete-time state-space form is stated as follows:

$$L[k+1] = \Phi\, L[k]k + \Gamma u[k],$$
$$y[k] = C^T L[k], \quad (1)$$

Where $p$ is a model order; $L[k] = \begin{bmatrix} l_1[k] & l_2[k] & \cdots & l_p[k] \end{bmatrix}^T$ is a state vector, consisting of Laguerre functions; $\Phi$ is $(p \times p)$ lower triangle matrix; $\Gamma$ is $(p \times 1)$ column vector:

$$\Phi = \begin{bmatrix} \psi & 0 & 0 & \cdots & 0 \\ \vartheta & \psi & 0 & \ddots & 0 \\ -\psi\vartheta & \vartheta & \psi & \ddots & \vdots \\ \vdots & \vdots & & \ddots & 0 \\ (-\psi)^{p-2}\vartheta & (-\psi)^{p-3}\vartheta & \cdots & \vartheta & \psi \end{bmatrix}, \quad \Gamma = \sqrt{\vartheta}\begin{bmatrix} 1 \\ -\psi \\ \psi^2 \\ \vdots \\ (-\psi)^{p-1} \end{bmatrix}, \quad (2)$$

$$C = \begin{bmatrix} c_1 & c_2 & \cdots & c_p \end{bmatrix}^T, \quad (3)$$

where $\psi$ is a scaling factor that should be in the range of $0 \leq \psi < 1$ to ensure system stability.

The task of Laguerre network parametrical identification of the order $p$ is reduced to determination components of vector $C$ (3) and scaling factor $\psi$.

Taking into account that discrete-time model (1) is used as the linear part of block-oriented models, it is necessary to adapt traditional structures taking into account this feature.

Modified Hammerstein model, which includes a static nonlinearity in the part describing the effect of input actions on the state vector value, can be represented in the following form:

$$L[k+1] = \Phi L[k] + \Gamma \Xi(u[k]),$$
$$y[k] = C^T L[k], \quad (4)$$

where $\Xi[k] = \begin{bmatrix} g_1(u[k]) & g_1(u[k]) & \cdots & g_m(u[k]) \end{bmatrix}^T - g_i:\ ^m \rightarrow\ $ is a nonlinear static functions; $L[k] \in\ ^n$ is a Laguerre model state vector; $u[k] \in\ ^m$ is an input vector.

Wiener model is stated in a similar way:

$$\begin{aligned} L[k+1] &= \Phi L[k] + \Gamma u[k], \\ y[k] &= \Psi\left(C^T L[k]\right), \end{aligned} \quad (5)$$

where $\Psi(\cdot): \mathbb{R}^n \to \mathbb{R}$ – static nonlinear function that connects the vector of Laguerre net state with output of block-oriented model.

As a result of combining of the two previous models (4) and (5) we receive Hammerstein-Wiener model in state-space form:

$$\begin{aligned} L[k+1] &= \Phi L[k] + \Gamma \Xi(u[k]), \\ y[k] &= \Psi(C^T L[k]). \end{aligned} \quad (6)$$

To evaluate the quality of the solution to the identification problem mean square error (MSE) is used in the following form:

$$MSE = \frac{1}{N} \sum_{i=1}^{N} \left(Y_i - \hat{Y}_i\right)^2, \quad (7)$$

where $Y_i$ – output values vector of the plant; $\hat{Y}_i$ – output values vector of the model, $N$ – the length of the set.

## 3. SIMULATION RESULTS

For the formulation of the cone crusher nonlinear dynamic model and its further parametric identification, the results of the MATLAB computational experiment based on plant model formulated as distributed parameter system [1], were used as input and output data. The capacity of the cone crusher for ore size classes -9.1+6.7 mm and below is accepted as output. The input is represented by the change closed size setting (CSS). Taking into account the significant persistence of the process, the sampling interval was accepted to be equal to 1 second, as a result test set amounted to 2000 samples.

The identification of Laguerre model (1) was made with using of recursive least squares (RLS) algorithm [7]. Based on received parameters, the vector of coefficients was chosen, which provides a better approximation of experimental data in terms of MSE. The compassion of reactions of the model and the plant to identical input effect is shown in Fig. 1. The graphs show that Laguerre model well follows the dynamics of the system, but cannot adequately display the nonlinear amplification coefficient.

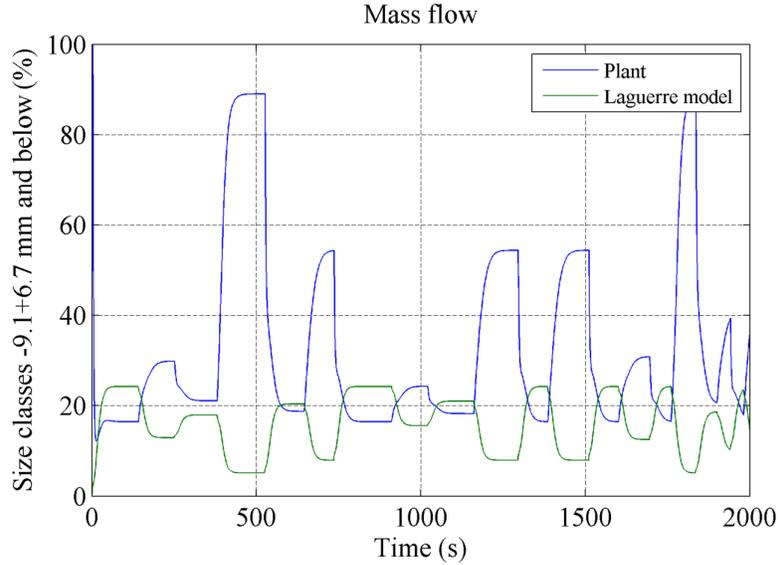

Figure 1. The comparison of the approximation efficiency of the cone crusher dynamics while usingLaguerre model with an adaptive algorithm of the parameter identification

Using the mathematical description of the block-oriented models (4-6), the dynamic characteristics approximation of the cone crusher was made by means of MATLAB. The evaluation of static nonlinearities on the input and output of models is made with using of the piecewise linear functions. At this the additive noise is applied on the plant output to simulate the impact of uncontrolled external disturbances, this noise is made by the sequence of random variables with a normal distribution that has the standard deviation σ.

Table 1. The comparison of the approximation efficiency of the dynamic characteristics of the plant at various amplitudes of disturbances

| The standard deviation, $\sigma$ | MSE | | | |
|---|---|---|---|---|
| | Hammerstein model | Wiener model | Hammerstein-Wienermodel | Hammerstein-Wienermodel with ARX |
| 0.01 | 20.72 | 87.89 | 7.06 | 15.12 |
| 0.05 | 20.73 | 93.33 | 7.13 | 15.44 |
| 0.1 | 20.78 | 58.24 | 7.25 | 16.36 |
| 0.25 | 20.81 | 61.15 | 7.43 | 18.23 |
| 0.5 | 20.98 | 86.61 | 8.89 | 23.97 |
| 1 | 20.77 | 75.18 | 10.65 | 29.68 |
| 5 | 44.84 | 100.6 | 30.61 | 45.07 |

The results given in Table 1 show, that the most accurate identification is achieved with the usage of Hammerstein-Wiener model with specification of parameter vector of Laguerre model with the adaptive RLS algorithm. Hammerstein model with analog algorithm of linear part parameters identification and Hammerstein-Wiener model with the linear block in the form of autoregressive (ARX) structure have worse, but acceptable readings. It should be noted that first two mentioned non-linear models show high stability of the identification process at the change of amplitudes of disturbances within range of $\{\sigma \in \ |0.01 \leq \sigma \leq 1\}$. The mean square deviation of the MSE criterion from the mean value for them is 0.0867 and 1.31. And for Hammerstein-Wiener model with ARX, this reading amounts to 3.4. The series of numerical experiments allowed to establish the instability of the latest model to the effects of uncontrolled high amplitude disturbances $\{\sigma \in \ | \sigma > 1\}$. At this the values of MSE vary within wide range from 45.81 to 468.23.

Comparative analysis showed that Wiener model cannot provide sufficient approximation accuracy of the dynamic characteristics of the plant. During the process it is unstable due to the high sensitivity to disturbances on the output. This feature is primarily related to the structural implementation of the model. Therefore, Wiener model is useful only in the control systems, which have additional devices to eliminate noise.

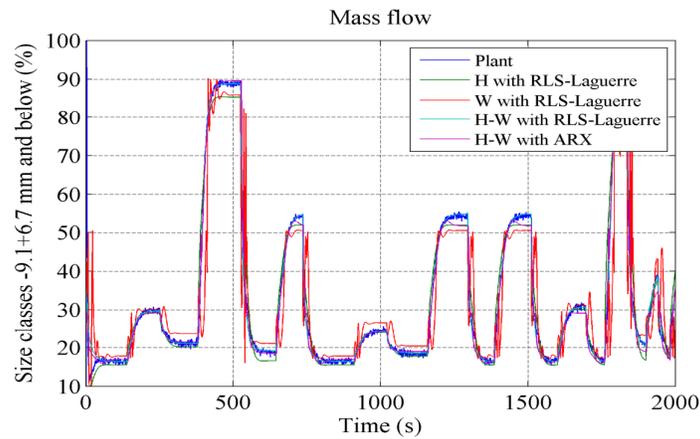

Figure 2. Comparison of identification efficiency of nonlinear models of cone crushers of different structure (σ = 0.5)

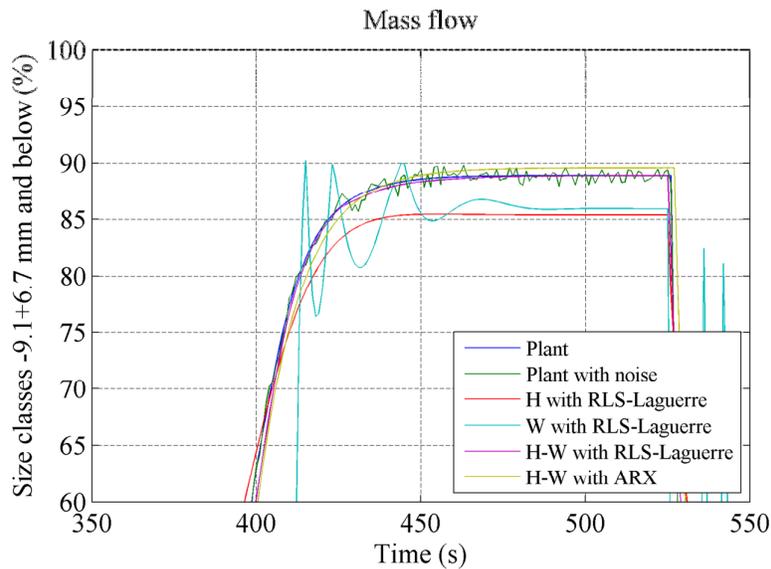

Figure 3. Comparison of identification efficiency of nonlinear models of cone crushers of different structure σ = 0.5 (enlarged fragment)

The temporal characteristics of the original plant and nonlinear models obtained in the result of identification are shown in Figures 2-6. For descriptive reasons the real output of the plant and output with the applied external additive disturbances are showed on enlarged graphs (Figures 3, 5, and 6).

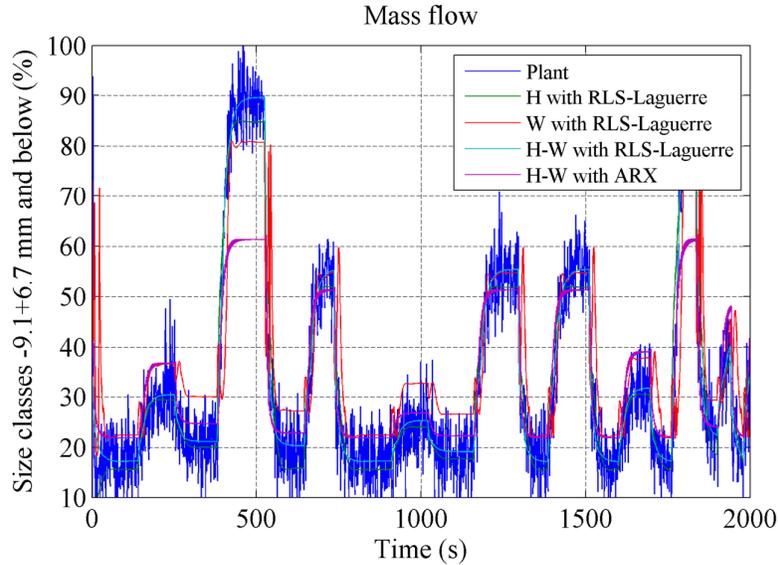

Figure 4. Comparison of identification efficiency of nonlinear models of cone crushers of different structure
$\sigma = 5$

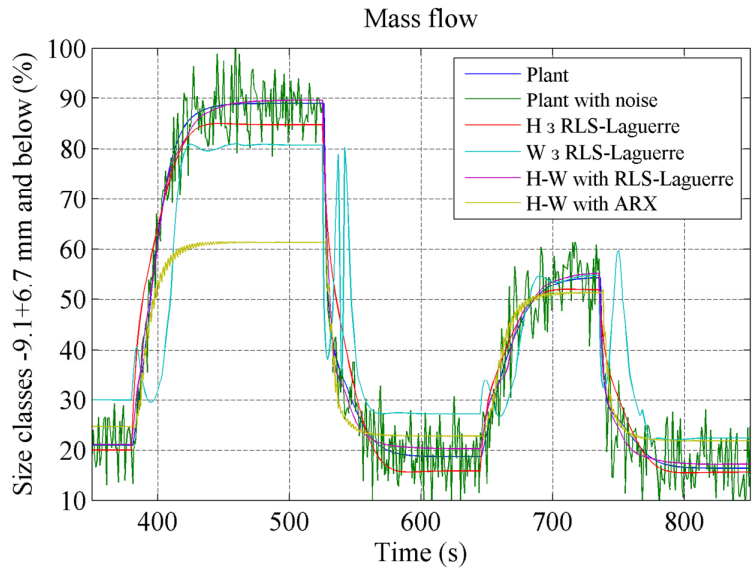

Figure 5. Comparison of identification efficiency of nonlinear models of cone crushers of different structure
$\sigma = 5$ (enlarged fragment)

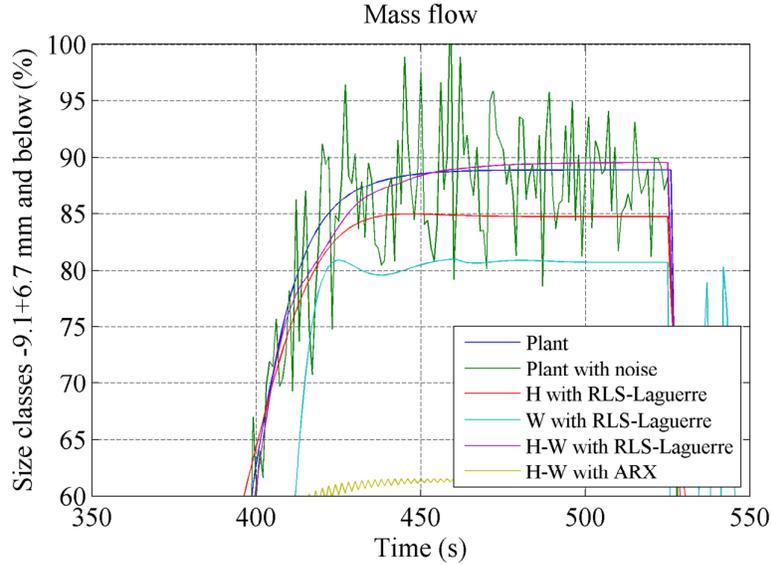

Figure 6. Comparison of identification efficiency of nonlinear models of cone crushers of different structure
$\sigma = 5$ (enlarged fragment)

The obtained results confirm high quality of approximation at using of Hammerstein-Wiener model with RLS algorithm of determination of the linear block parameters vector, the results also show instability of Wiener model.

Another significant advantage of using of the structure, which showed the best quality of identification is that it functions as a digital filter with a high level of reconstruction of the original signal without noise.

## 4. CONCLUSIONS

The studies of using of Laguerre model with an adaptive recursive least squares algorithm identification of its parameters as linear part of Hammerstein, Wiener and Hammerstein-Wiener models showed that Hammerstein-Wiener model has the best approximation of temporal characteristics of the model to plant dynamics under effect of uncontrolled disturbances. Its accuracy is by 64.3% better than the accuracy of Hammerstein model and is by 59.24%better than the one of Hammerstein-Wiener model with a linear block based on autoregressive structure (ARX) at change of the amplitude ofadditivedisturbances within $\{\sigma \in \mathbb{R} \mid 0.01 \leq \sigma \leq 1\}$. Wiener model is unsuitable for operation in the conditions of disturbances of random amplitude due to its sensitivity to the value of the plant output. Hammerstein-Wiener model with ARX loses its stability at $\{\sigma \in \mathbb{R} \mid \sigma > 1\}$.

**Authors**

**MykhailenkoOleksii** Received a master's degree inelectromechanics at KrivyyRih technical University, KrivyyRih, Ukraine, in 2008. Current research interests include the development of identification algorithms for non-linear models, digital filtering, predictive control of technological processes, development of digital control systems. 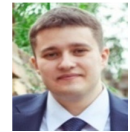